\documentclass{emulateapj}
\usepackage{apjfonts}
\usepackage{amsmath}
\usepackage{multirow}
\usepackage{url}
\shorttitle{GeV emission from Arp 220}
\begin{document}

\title{{First detection of GeV emission from an ultraluminous infrared galaxy: Arp 220 as seen with the Fermi Large Area Telescope}}

\author{Fang-Kun Peng\altaffilmark{1,2}, Xiang-Yu
Wang\altaffilmark{1,2}, Ruo-Yu Liu\altaffilmark{3}, Qing-Wen
Tang\altaffilmark{4},  Jun-Feng Wang\altaffilmark{5}}
\altaffiltext{1}{School of Astronomy and Space Science, Nanjing
University, Nanjing 210093, China; xywang@nju.edu.cn}
\altaffiltext{2}{Key laboratory of Modern Astronomy and
Astrophysics (Nanjing University), Ministry of Education, Nanjing
210093, China} \altaffiltext{3}{Max-Planck-Institut f\"ur
Kernphysik, 69117 Heidelberg, Germany} \altaffiltext{4}{School of
Science, Nanchang University, Nanchang 330031, China}
\altaffiltext{5}{Department of Astronomy and Institute of
Theoretical Physics and Astrophysics, Xiamen University, Xiamen,
Fujian 361005, China}

\begin{abstract}
Cosmic rays (CRs)  in starburst galaxies  produce high energy
gamma-rays by colliding with the dense interstellar medium (ISM).
Arp 220  is  the nearest  ultra luminous infrared  galaxy (ULIRG)
that has star-formation at extreme levels, so it has long been
predicted to emit high-energy gamma-rays. However,  no evidence of
gamma-ray emission was found despite intense efforts of search.
Here we report the discovery of high-energy gamma-ray emission
above 200 MeV from Arp 220 at a confidence level of $\sim 6.3
\sigma $ using 7.5 years of \textsl {Fermi} Large Area Telescope
observations. The gamma-ray emission shows no significant
variability over the observation period and it is consistent with
the quasi-linear scaling relation between the gamma-ray luminosity
and total infrared luminosity for star-forming galaxies,
suggesting that these gamma-rays arise from CR interactions. {As
the high density medium of Arp 220 makes it} an ideal CR
calorimeter, the gamma-ray luminosity can be used to measure the
efficiency of powering CRs by supernova (SN) remnants  given a
known supernova rate in Arp 220. We find that this efficiency is
about $4.2\pm2.6\%$ for CRs above 1 GeV.

\end{abstract}


\section{Introduction}
Nearby star-forming and starburst galaxies have been identified to
be GeV to TeV gamma-ray sources \citep{2009Sci...326.1080A,2009Natur.462..770V,
2010ApJ...709L.152A,2012ApJ...755..164A}. Cosmic rays (CRs)
accelerated by supernova remnants therein interact with the
interstellar gas and produce neutral pions (schematically written
as $p+p\rightarrow \pi^0+$ other products), which in turn decay
into high-energy gamma-rays ($\pi^0\rightarrow \gamma+\gamma$).
High energy gamma-ray emission is thus a powerful tool to diagnose
the non-thermal emission related with CRs in star-forming
galaxies.

Arp 220, at a distance of $d=77$ Mpc, is the nearest
ultra-luminous infrared galaxy (ULIRG) with a total infrared
luminosity of $(1-2)\times10^{12} L_{\odot}$
\citep{2003AJ....126.1607S,2004ApJ...606..271G,2011ApJ...743...94R}.
The system is a merger of two galaxies and contains two dense
nuclei separated by 350 pc. {Both nuclei have high star formation
rate and dense molecular gas.} Arp 220 has a molecular gas mass of
$10^{10}M_\odot$ \citep{1997ApJ...484..702S} and a total star
formation rate of $240\pm 30 \rm M_{\odot} /{\rm yr}$, calculated
with the far-infrared luminosity \citep{2003MNRAS.343..585F}.
Radio detection of supernovae suggests a high rate of $4 \pm
2/{\rm yr}$
\citep{1998ApJ...493L..17S,2006ApJ...647..185L,2007ApJ...659..314P}.
Although Arp 220 could contain active galactic nuclei (AGNs)
\citep{2013arXiv1303.2630P,2015ApJ...814...56T},  the observed
high supernova rate indicates that star formation provides a
substantial fraction of the power radiated by the nuclei.

As a huge reservoir of molecular gas, the nuclei of Arp 220 should
be complete proton calorimeters, i.e. the  CR protons lose all of
their energy to secondary pions via collisions with the gases
before escaping
\citep{2004ApJ...617..966T,Thompson2007,2011ApJ...734..107L,2015MNRAS.453..222Y}.
Given a high CR emissivity and a high efficiency of converting CR
energy  into gamma-rays, Arp 220 was predicted to emit GeV to TeV
gamma-rays
\citep{2004ApJ...617..966T,2011ApJ...734..107L,2015MNRAS.453..222Y}.
There have been intense efforts to search for  GeV emission from
Arp 220
\citep{2011ApJ...734..107L,2012ApJ...755..164A,2014ApJ...794...26T}.
\citet{2014ApJ...794...26T} obtained an upper limit $F_{\rm
0.1-100 GeV} < 3.5 \times 10^{-9}$ ph cm$^{-2}$ s$^{-1}$ at 95\%
confidence level by using 68 months of \textsl{Fermi}/LAT data.
Arp 220 was also observed by the VERITAS telescopes at TeV
energies for more than 30 hours, but no significant excess over
the background is found \citep{2015arXiv150805807F}.  Here we
report the  detection of GeV emission from Arp 220 using 7.5 years
of \textsl{Fermi}/LAT data reprocessed with Pass 8, the newest
event-level analysis, which significantly improves the acceptance
and angular resolution of the
instrument\footnote{\url{http://fermi.gsfc.nasa.gov/ssc/}}.

We present a description of the analysis and show the results in
\S 2. Then we discuss the origin of the $\gamma $-ray emission (\S
3) and use it to constrain the cosmic ray acceleration efficiency
(\S 4). Finally we give the conclusions in \S 5.

\section{Data analysis}
\subsection{Data selection}
We choose the \textsl{Fermi}/LAT
survey data with P8R2 SOURCE event class from 2008-08-04 to
2016-01-22, and use the current \textsl{Fermi} science tools
version v10r0p5 for data analysis \citep{2009ApJ...697.1071A}. The maximum zenith angle of $< 90^{\circ}$ is
selected to reduce the Earth Limb emission. The data are filtered
with the recommended cuts (DATA\_QUAL $> 1$) \&\& (LAT\_CONFIG ==
1) \&\& ABS(ROCK\_ANGLE)$< 52^{\circ}$. Photons with energy range
between 200 MeV and 100 GeV are taken into consideration. The region
of interest (ROI) is a circle with a radius of $10^{\circ}$ centered at the position of Arp 220 (i.e., ${\rm
RA=233.738^{\circ}, Dec=23.5032^{\circ}}$, J2000).
We use the FRONT+BACK converting photons. The
3FGL sources \citep{2015ApJS..218...23A} within
$15^{\circ}$ of the ROI center are included in the source
model. The galactic diffuse emission and isotropic emission are modeled with
"gll\_iem\_v06.fits" and "iso\_P8R2\_SOURCE\_V6\_v06.txt"
respectively. We also consider the CLEAN data and binned method to
check our results and no statistical difference is found.

\subsection{Spatial and Spectral Analysis}
We start with a preliminary unbinned likelihood analysis on the
ROI with the tool \emph{gtlike}, only including
"gll\_iem\_v06.fits", "iso\_P8R2\_SOURCE\_V6\_v06.txt" and point
sources of 3FGL within $15^{\circ}$ from the ROI center in the
source model file. The residual test-statistic (TS) under this
scenario is shown in the left panel of Fig. \ref{tsmaps}, which
shows a gamma-ray  excess  around  Arp 220. We add one point
source with a power-law spectrum at the local maxima of the TS map
in the source model file. We re-fit the data and find the best-fit
$\gamma $-ray emission position with the tool \emph{gtfindsrc}. We
then change the position of the point source to the obtained
best-fit location of $\gamma$-ray emission in the source model
file. Repeating the standard analysis threads and computing the TS
map, although getting a significantly improved fit, {we note that
there is still some residual emission to the northwest of the ROI
center.} In order to explore the high energy emission more
accurately near Arp 220  and reduce the contamination from nearby
background sources,  a second point source is added to the source
model based on the residual $\gamma $-ray emission. By utilizing
the tool \emph{gtfindsrc}, we optimize the position of these two
point source, namely, P1 ($233.677^{\circ}, 23.5163^{\circ})\pm
0.054^{\circ}$ which is near  Arp 220 and P2 ($233.239^{\circ},
23.8049^{\circ})\pm 0.168^{\circ}$ which locates at a position
about $0.5^\circ$ away from  Arp 220, as marked in Fig
\ref{tsmaps}. Next, we re-fit the data with point sources P1 and
P2 in the source model file, and get a better result with $\rm
\Delta \ Likelihood = 8$ {(i.e. we get a global improved fitting
with significance $\rm TS = 16$)}, in comparison with that only
one point source is added into the source model file. The
significance of $\rm TS_{P2}=22$ suggests that P2 should be a
genuine additional $\gamma$-ray source. The bottom panel in Fig.
\ref{tsmaps} illustrates that there is little residual
$\gamma$-ray emission near the vicinity of Arp 220 region. The
source model file including point source P1 and P2 will be used in
all the subsequent analysis.

We now check possible alternative candidates for P1 and P2 in  the
following catalogs: CRATES Flat-Spectrum Radio Source Catalog
\citep{2007ApJS..171...61H}, Veron Catalog of Quasars \& AGN, 13th
Edition \citep{2010A&A...518A..10V} and Candidate Gamma-Ray Blazar
Survey Source Catalog \citep{2008ApJS..175...97H}. No other
sources are found in the vicinity of $r_{95}$ of P1 except for Arp
220. With only a separation of $0.058^{\circ}$ between  P1 and Arp
220, it is reasonable to ascribe the high energy emission from P1
to Arp 220. By contrast, the separation between P2 and Arp 220
reaches $0.547^\circ$ and there are four candidates around
$r_{95}$ of P2 in the above three catalogs: CRATES J153246+234400
($4.902'$), a FSRQ with unknown redshift; SDSS J15323+2345
($9.080'$), a QSO with redshfit $z=1.465$; SDSS J15337+2358
($14.643'$), a Seyfert 1 galaxy with $z=0.067$; and 1WGA
J1533.8+2356 ($15.315'$), an AGN of unknown type located at
$z=0.232$. Values in the parentheses are the angular offsets of
these sources from P2. We suspect that the gamma-ray emission from
P2 is contributed by one or more of these AGNs.

We assume power-law spectra in the range of 0.2-100 GeV for both
P1 and P2 in the aforementioned best source model files. The
best-fit photon index for emissions of P1 is $\Gamma_1=2.35\pm
0.16$ with a total flux of $F_1=1.76\pm 0.52 \times 10^{-9}$ ph
cm$^{-2}$ s$^{-1}$. Regarding P2, the photon index is
$\Gamma_2=2.45\pm 0.19$ and the total flux is $F_2=1.45\pm 0.52
\times 10^{-9}$ ph cm$^{-2}$ s$^{-1}$. The corresponding energy
flux in 0.2-100 GeV are $1.92\pm 0.43 \times 10^{-12}$ erg
cm$^{-2}$ s$^{-1}$ and $1.39\pm 0.40 \times 10^{-12}$ erg
cm$^{-2}$ s$^{-1}$ for P1 and P2, respectively. The spatial and
spectral results of P1 and P2 are summarized in Table \ref{para}.
The spectral energy distribution (SED) of P1 (Arp 220) is shown in
Fig. \ref{lcsed}. We do not discuss P2 further in this paper since
the topic is on the gamma-ray emission from Arp 220.

\section{Origin of the gamma-ray emission from Arp 220 }
Gamma-ray emissions in star-forming galaxies are usually thought
to originate from cosmic  ray protons via collision with the gases
inside the galaxies. {This may be especially the case with Arp 220
as it has a high supernova rate and high gas density.}
Alternatively,  Arp 220 could have  hidden
AGNs\citep{2013arXiv1303.2630P,2015ApJ...814...56T},  which in
principle could  give rise to gamma-ray emission as well.
CR-induced gamma-ray emission in a star-forming galaxy is expected
to be stable, while gamma-ray emission induced by AGNs should be
temporally variable \citep{2015arXiv150805301G}. Thus, we examine
the  flux variability of Arp 220 (P1). We create a set of time
bins for the light curve of photons with energy $>400\,$MeV. In
the first trial, the full observation period is divided linearly
into 5 equal time bins. Each time bin is fitted by a separate
maximum likelihood analysis, resulting in each bin with detection
significance $\rm TS > 4$. We further check a finer light curve
with 12 time bins. The $\chi^2$ goodness-of-fit test is consistent
with a constant flux with a reduced $\chi^2$ of 0.83 for data
points with detection significance $\rm TS>1$. The results is
shown in Fig \ref{lcsed}. None of the two cases show significant
changes in flux over the period of LAT observations \footnote{{The
gamma-ray excess indicated by the P7 data in Tang et al. (2014)
locates at a position within $r_{95}$ of P2, so this excess has
nothing to do with Arp 220 (P1). The source of this gamma-ray
excess in Tang et al. (2014) was likely FSRQ CRATES J
153246+234400, which shows variability. }}. Generally speaking,
the temporal behavior of GeV emission from Arp 220 resembles that
of other starburst galaxies, such as M82 and NGC 253.

Meanwhile, there is a clear positive empirical relation between
the $\gamma$-ray luminosity $L_{\rm 0.1-100 GeV}$ and total
infrared luminosity $L_{\rm 8-1000 \mu m}$ in local group galaxies
and nearby star-forming galaxies{\footnote{Although the origin of
high energy emission of  two starbursting Seyfert 2 galaxies NGC
1068 and NGC 4945 are not yet definitely established, {lack of
$\gamma$-ray variability and the consistent behavior with this
correlation  favor a CR origin.}}\citep{2012ApJ...755..164A}. With
a pure sample of star-forming galaxies,
\citet{2014ApJ...794...26T} have drawn a similar relation and
extended the relation to a luminous infrared galaxy NGC 2146. The
best simple linear fit for nearby star-forming galaxies excluding
Arp 220 gives a relation of
\begin{equation}
{\rm log}\bigg{(}\frac{L_{\rm 0.1-100 GeV}}{\rm erg \ s^{-1}}\bigg{)}= \alpha +\beta {\rm log}\bigg{(}\frac{L_{\rm \rm 8-1000 \mu m}}{\rm erg \ s^{-1}}\bigg{)}.
\end{equation}
(see Fig. \ref{relations}), where $\alpha =-14.98\pm 4.87$ and
$\beta=1.24\pm 0.11$ are the intercept and slope respectively. The
Pearson's correlation coefficient is 0.98 with a chance
probability of $p \sim 10^{-4}$.  With a luminosity of $L_{\rm
0.1-100 GeV}=1.78\pm 0.30 \times 10^{42}$ erg s$^{-1}$, Arp 220 is
consistent with this quasi-linear relation. It is also consistent
with the scaling relation between the gamma-ray luminosity and
radio luminosity found for star-forming galaxies
\citep{2012ApJ...755..164A}. The fact that the gamma-ray emission
has no significant variability over the observation period and it
is consistent with the above scaling relation suggests that the
gamma-ray emission is likely to arise from CR interactions.

\section{Measuring the CR acceleration efficiency}
As the ejecta of a SN encounters the ambient ISM,  a strong shock
will be generated and a fraction of the kinetic energy of the
ejecta will be transferred into cosmic rays. The gamma-ray
luminosity of a galaxy depends on both the efficiency of
converting SN kinetic energy to cosmic rays and the efficiency of
converting CR energy to gamma rays (i.e., the proton-proton
collision efficiency). {Because of the very dense ISM in Arp 220,
it has been widely suggested that Arp 220 is an ideal proton
calorimeter, since the collisional energy loss time of CR protons
is much shorter than the  time that CRs escape out of the galaxy}
\citep{2011ApJ...734..107L,2015MNRAS.453..222Y}. Then its
gamma-ray luminosity provides a direct tool to measure the
efficiency that SN explosion power goes into high-energy CRs.

Assuming that SNRs are the sole source of CR protons,  and that a
constant fraction  ($\eta$) of SN kinetic energy  is transferred
into CR protons with energy above 1 GeV, the total injected CR
power reads
\begin{equation}\label{Lcr1}
L_{\rm CR }({\rm >1GeV})=1.3\times 10^{44} {\rm erg\ s^{-1}}
E_{51} \eta \left(\frac{\Gamma_ {\rm SN}}{4 \,{\rm
yr^{-1}}}\right),
\end{equation}
where $E$ is the SN kinetic energy, which is typically
$10^{51}{\rm erg}$, and $\Gamma_{\rm SN}=4\pm 2/\rm yr $ is the SN
rate obtained by the radio observations
\citep{1998ApJ...493L..17S,2006ApJ...647..185L,2007ApJ...659..314P}.
On the other hand,  by assuming that the system reaches a steady
state, one can derive  the injected CR power from the observed
gamma-ray luminosity. As leptonic emission is expected to become
increasingly important at lower energies, contaminating the
estimate of the power in CR protons,  we only consider the
$\gamma$-ray emission above 1 GeV.  Then we have
\begin{equation}\label{Lcr2}
L_{\rm CR}(>1{\rm GeV})=3L_{\gamma}(>1\rm
GeV)(\Gamma-1)\beta_\pi^{-1}.
\end{equation}
Here the prefactor 3 is due to that only 1/3 of the lost CR power
goes into gamma-rays, as inelastic proton-proton collisions
produce both neutral and charged pions. The factor $\Gamma-1$
arises from the fact that a fraction $(\Gamma-2)/(\Gamma-1)$ of
the energy of CRs above 1 GeV is transferred to lower energy CRs,
where $\Gamma$ is the CR spectral index, which is equal to the
photon index of the gamma-ray emission in our case. Since not all
the pionic gamma rays produced by CRs will have energies above
1\,GeV, we need another correction factor $\beta_\pi\simeq0.6$
(for $\Gamma=-2.35$) to account for this effect
\citep{2011ApJ...734..107L}. Given that the luminosity of Arp 220
above 1 GeV is $L_{\gamma}(\rm
>1GeV)= (0.75\pm 0.28)\times 10^{42}$ erg s$^{-1}$, we can
obtain the CR acceleration efficiency from Eqs.~\ref{Lcr1} and
\ref{Lcr2}, i.e.
\begin{equation}
\eta \simeq (4.2\pm 2.6)\% E_{51}^{-1}
\left(\frac{\beta_{\pi}}{0.6}\right)^{-1} \left(\frac{\Gamma_ {\rm
SN}}{4 {\rm yr^{-1}}}\right)^{-1}.
\end{equation}
Note that here only the  uncertainties in $\Gamma_{\rm SN}$ and
$L_{\gamma}(\rm >1GeV)$ are  taken into consideration to estimate
the uncertainty in $\eta$. To our knowledge, this is the first
time that such an efficiency is obtained  directly from the
gamma-ray observations. {The obtained efficiency in Arp 220 is
consistent with estimates of a 3-10\% efficiency  in the Milky Way
\citep{2010ApJ...722L..58S} and a 6\% efficiency in NGC 253 and
M82 \citep{2012ApJ...755..106P}.}

\section{Conclusions}
Through analysis of 7.5 years of \textsl {Fermi}/LAT observations,
we find high-energy gamma-ray emission above 200 MeV from Arp 220
with a detection significance about $6.3 \sigma$. This is the
first time detection of GeV emission from an ULIRG.  The
reconstructed energy spectrum is best modeled by a single power
law with photons index $\Gamma=2.35\pm 0.16$ and the integrated
flux in 0.2-100 GeV is $F_{0.2-100 \rm GeV} =(1.76\pm 0.52) \times
10^{-9}$ ph cm$^{-2}$ s$^{-1}$. The fact that no evidence of
significant variability is found and that the gamma-ray emission
is consistent with some scaling relations  for star-forming
galaxies implies that the gamma-ray emission of Arp 220 is likely
from CR-induced diffuse emission. As Arp 220 is an ideal proton
calorimeter, the gamma-ray luminosity can be used  to directly
measure the CR acceleration efficiency by SNRs, which gives an
efficiency of $4.2\pm2.6\%$ for CRs above 1 GeV.

\section*{Acknowledgments}
We thank Anton Prosekin and Gwenael Giacinti for useful
discussions and the anonymous for a helpful report. This work has
made use of data and software provided by the \textsl{Fermi}
Science Support Center. It has also made use of the NASA/IPAC
Extragalactic Database (NED) which is operated by the Jet
Propulsion Laboratory, California Institute of Technology, under
contract with the National Aeronautics and Space Administration.
This work is supported by the 973 program under grant
2014CB845800, the NSFC under grants 11273016 and 11033002, and the
Excellent Youth Foundation of Jiangsu Province (BK2012011). J.W.
acknowledges support from NSFC grant 11473021 and the Fundamental
Research Funds for the Central Universities under grants
20720150168 and 20720160023.

\clearpage

\begin{figure}
\centering
\includegraphics[angle=0,width=10cm,height=8.5cm]{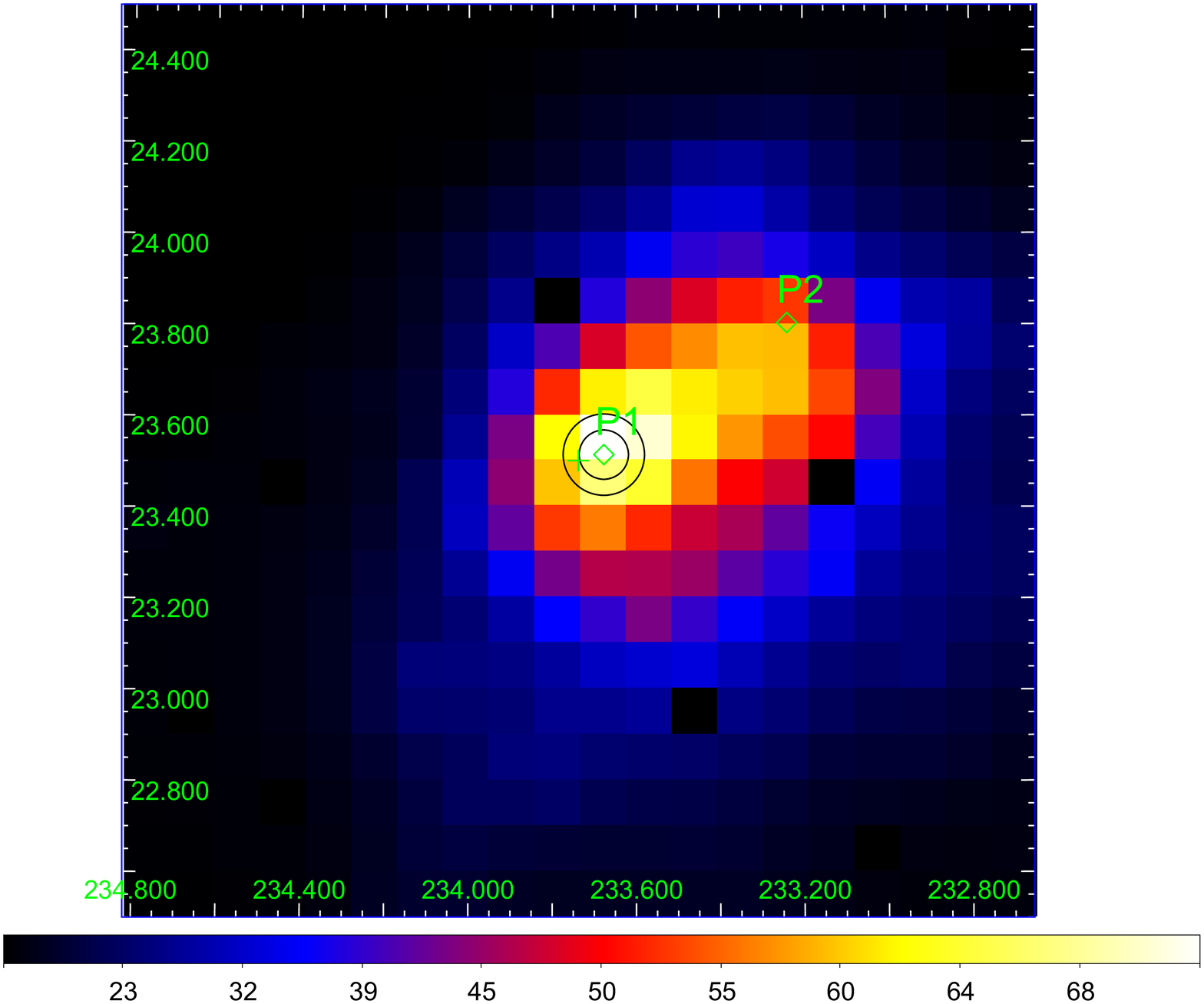}\\
\includegraphics[angle=0,width=10cm,height=8.5cm]{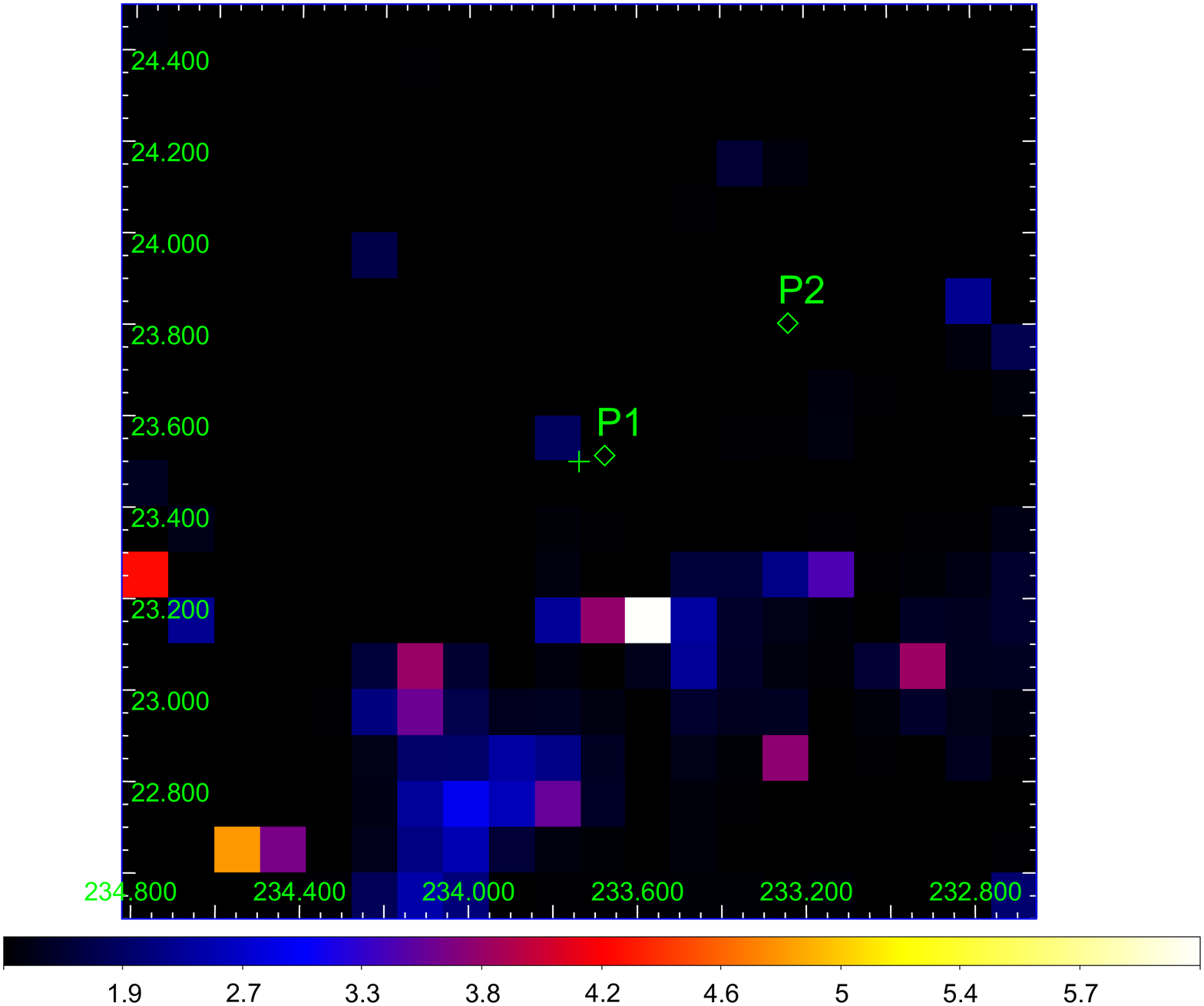}
\caption{TS map with binsize=$0.1^{\circ}$ above 200 MeV for the
$2^{\circ}\times 2^{\circ}$ region  centered at Arp 220 $\rm
(R.A., DEC.=233.738^{\circ}, 23.5032^{\circ})$. The green cross
denotes the position of Arp 220, and the black lines show the
1$\sigma$ and 2$\sigma$ errors around P1. The green diamonds
denote the positions of P1 and P2. The upper panel shows the
residual TS map after removing the emission from 3FGL point
sources, the galactic diffuse emission and isotropic emission. The
bottom panel shows the residual TS map after removing emissions
further from P1 and P2.} \label{tsmaps}
\end{figure}

\begin{figure}
\centering
\includegraphics[angle=0,width=10cm,height=5cm]{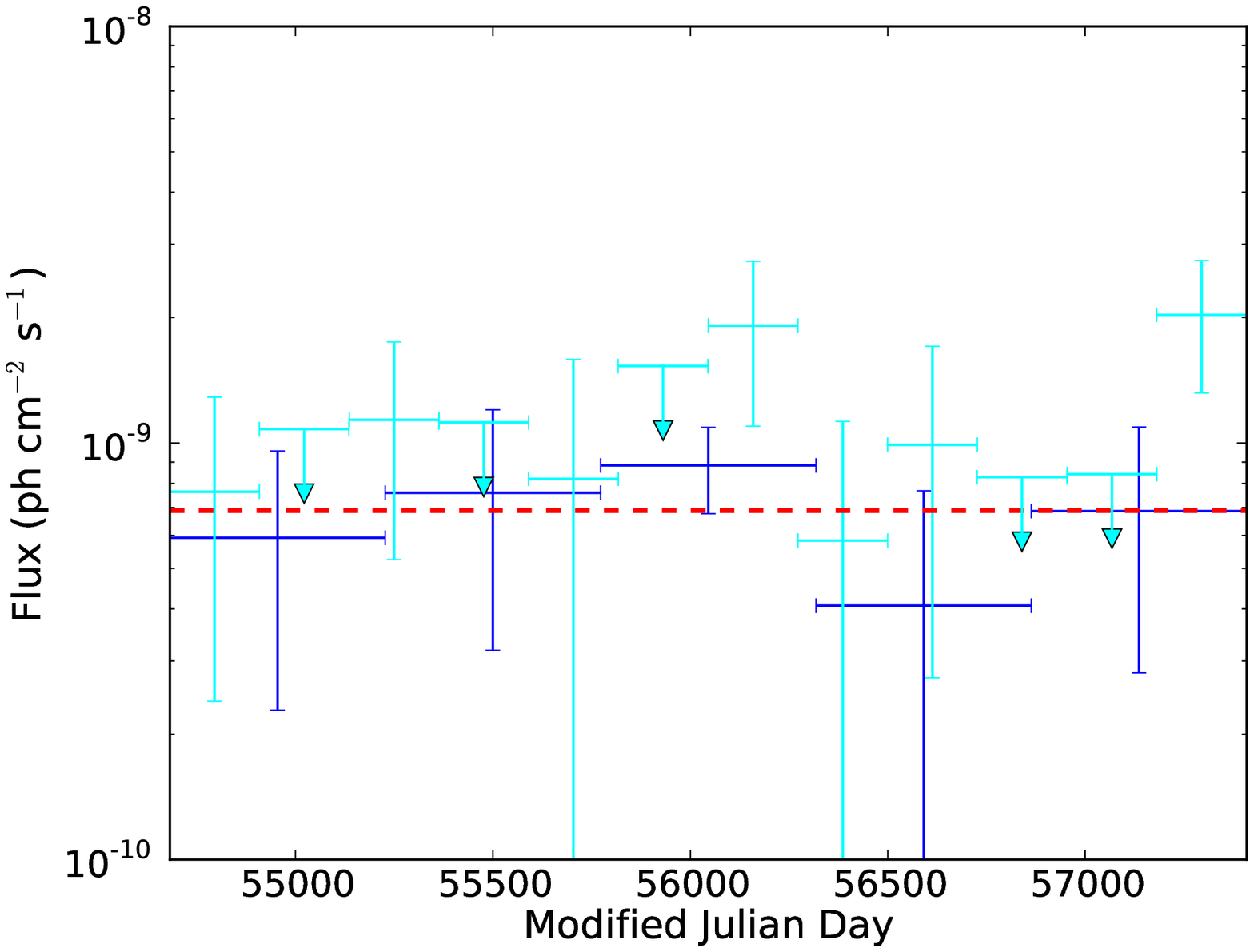}\\
\includegraphics[angle=0,width=10cm,height=5cm]{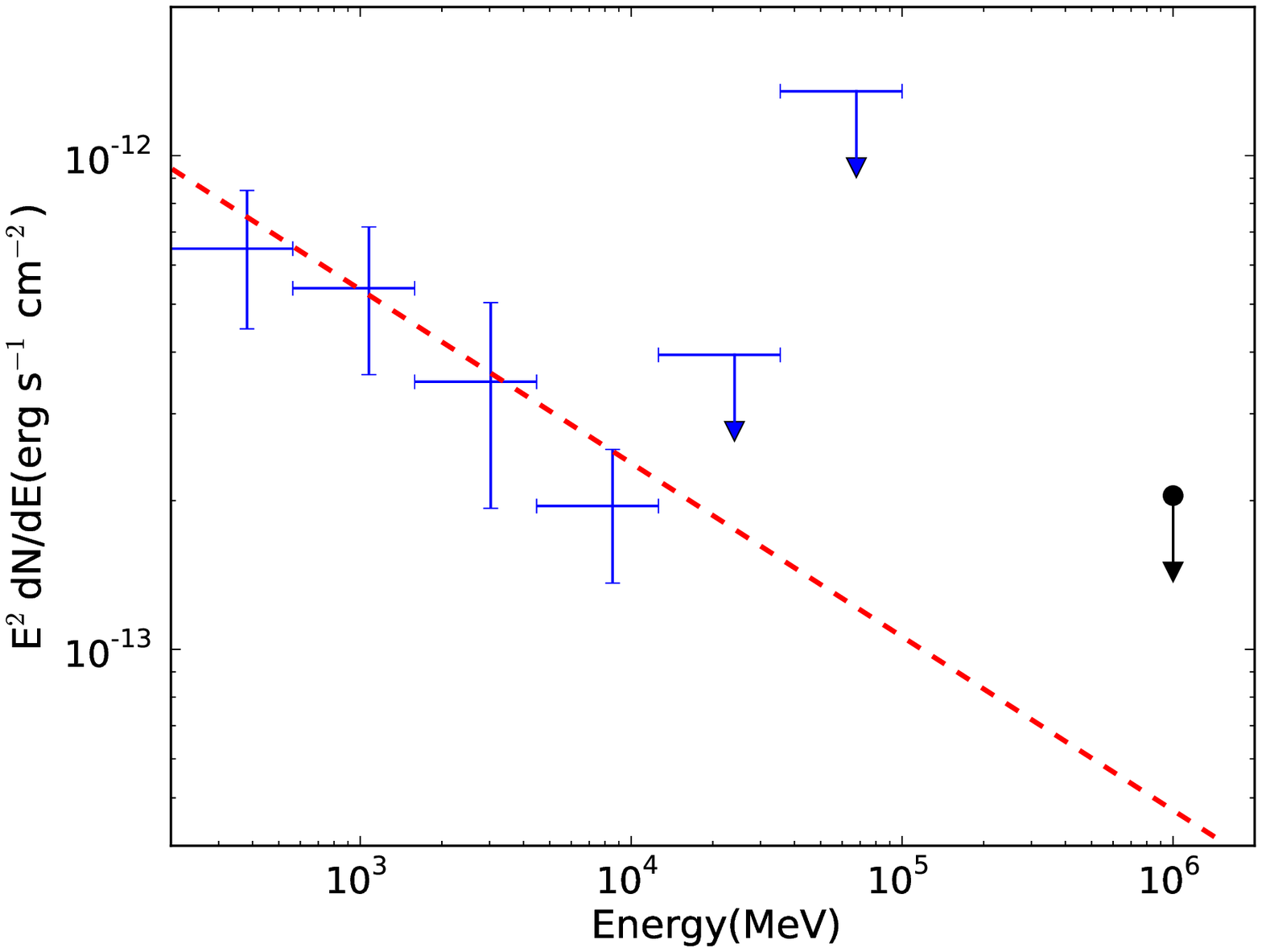}
\caption{Upper: \textsl{Fermi}/LAT (0.4-100 GeV) light curves of
P1 (Arp 220). The blue (cyan) data points indicate the light curve
for five (twelve) time bins. The  red dashed line illustrates the
maximum likelihood flux level for the $\sim 7.5$ years
observations; Bottom: The spectral energy distribution of P1 (Arp
220).  Upper limits of two blue data points  are at 95\%
confidence level. The red dashed line illustrates the best-fit
power-law spectral model in 0.2-100 GeV with the unbinned
analysis. The black data point is derived from the integral upper
limit flux observed by VERITAS \citep{2015arXiv150805807F}.}
\label{lcsed}
\end{figure}

\begin{figure}
\centering
\includegraphics[angle=0,scale=0.40]{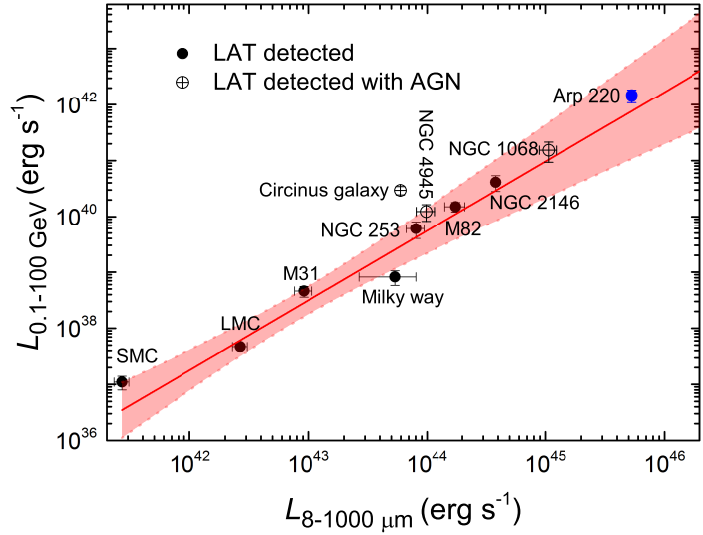}
\caption{Gamma-ray luminosity (0.1-100 GeV) vs. total infrared
luminosity (8-1000 $\rm \mu m$) for LAT-detected star-forming
galaxies and Seyferts. The red solid line is the best-fit
power-law relation for star-forming galaxies only (black filled
circles). Two dotted lines represent its 95\% confidence level
region around the best fit. Arp 220, marked with blue filled
circle, lies on the extrapolation of this relation to higher
luminosity. The total infrared luminosity data are taken from
\citet{2004ApJ...606..271G}, and $\gamma $-ray luminosities are
taken from \citet{2012ApJ...755..164A} and
\citet{2014ApJ...794...26T}. } \label{relations}
\end{figure}

\begin{table}
\small
\centering \caption{The best-fit spectral parameters of the point
sources around  Arp 220 region for energy band 0.2-100 GeV. The
fourth column is the angle distance between the best-fit position
of $\gamma $-ray excess and the position of Arp 220.}
\begin{tabular}{ccccccccc}
\hline \hline
 Point& Position & $r_{95}$ & Separation & Photon Flux & Energy Flux & $\Gamma$ &TS & Association \\
 & deg & deg & deg & $10^{-9}$ ph cm$^{-2}$s$^{-1}$ & $10^{-12}$ erg cm$^{-2}$s$^{-1}$ & &  &  \\
\hline
P1 &  (233.677, 23.5163)  &  0.090 & 0.058 & $1.76\pm0.52$ & $1.92\pm0.43$ &  $2.35\pm0.16$& 40& Arp 220   \\
P2 &  (233.239, 23.8049)  &  0.279 & 0.547 & $1.45\pm0.52$ &
$1.39\pm0.40$ &  $2.45\pm0.19$& 22&           \\
\hline
\end{tabular}
\label{para}
\end{table}

\end{document}